\documentstyle[twoside,fleqn,espcrc2]{article}
\title{Bose Condensates in TOP Traps Exhibit Circulating Superfluid Flows}

\author{Juhao Wu and A. Widom  
\address{Physics Department, Northeastern University, Boston MA 02115}
}

\begin{document}

\begin{abstract}
For spin one atoms localized in a quadrapole magnetic field gradient, 
the atoms may be impeded from spin flipping their way out from the 
center of the trap by the application of a rotating uniform magnetic 
field. From a quantum mechanical viewpoint, such a trap for a Bose 
condensate is equivalent to having a superfluid in a rotating bucket. 
Vorticity is then expected to be induced in the condensate fluid flow 
without the application of any further external perturbations.  
\end{abstract}

\maketitle

The two fluid model\cite{1} has been quite successful for describing 
liquid $He^4$ below the lambda temperature. The model asserts that the 
normal fluid component can easily carry vorticity, but the superfluid 
component can undergo only potential flows with velocity 
${\bf v}_s={\bf \nabla}\Phi $. However, it is presently understood 
that when superfluid $He^4$ is located in a rotating container, vorticity 
can enter into the superfluid flow in the form of ``vortex lines''
\cite{2,3} with a circulation $\kappa =\oint {\bf v}_s\cdot d{\bf r}$ 
quantized in units of $\kappa_0=(2\pi \hbar /M)$. 

In the light of recent progress in preparing Bose condensates in magnetic 
traps\cite{4,5,6}, there has been considerable interest in whether or not 
vorticity can play an important role in the experimental features of a 
mesoscopic superfluid. Our purpose is to point out that superfluid 
vorticity {\em must already be present} in those experiments which 
employ rapidly rotating magnetic fields, i.e. in the so-called ``TOP trap''
experiments. The detailed description of TOP traps will be reviewed in the 
work which follows. Here, we note that the implications of rotational 
vorticity for TOP trap experiments have {\em not} been previously explored.

Consider (at first) the quantum dynamic behavior of (say) superfluid $He^4$ 
in an arbitrarily shaped rotating bucket. Let $H$ denote the Hamiltonian of 
the superfluid in the bucket {\em if the bucket were not rotating}. 
The Hamiltonian $H$ does not depend on time. However, since the bucket 
does rotate at angular velocity 
\begin{equation}
{\bf \Omega}={\bf n}\big(d\theta /dt\big),
\end{equation}
where ${\bf n}$ is a unit vector along the axis of rotation, the 
fluid laboratory frame Hamiltonian actually develops a time dependence 
given by 
\begin{equation}
H(t)=S(t)HS^\dagger (t),
\end{equation}
\begin{equation}
S(t)=\exp\big(-i{\bf n\cdot L}\theta (t)/\hbar \big),
\end{equation}
where ${\bf L}$ is the total angular momentum of the fluid. 
All that is required to understand the above time dependent Hamiltonian 
$H(t)$ for a fluid in a rotating bucket is that 
the angular momentum of the fluid is the generator of rotations\cite{7}. 

A crucial step is to {\em eliminate the time dependence} from the 
Hamiltonian $H(t)$ by using the following quantum mechanical canonical 
transformation ($H(t)\to {\cal H}$),
\begin{equation}
{\cal H}=S^\dagger (t)H(t)S(t)-
i\hbar S^\dagger (t)\big(\partial S(t)/ \partial t\big),
\end{equation}
to a frame rotating along with the bucket. Employing Eqs.(1)-(4),  
one finds  
\begin{equation}
{\cal H}=H-{\bf \Omega \cdot L}.
\end{equation}
Note that the effective Hamiltonian ${\cal H}$ in Eq.(5) does 
{\em not} depend on time if the angular velocity ${\bf \Omega}$ does
not depend on time. We are thereby perfectly justified\cite{8} in using 
${\cal H}$ in Eq.(5) to describe a thermal equilibrium  Bose fluid  
within a rotating bucket; The Gibbs canonical distribution is   
\begin{equation}
\rho_{eq}=\exp\big((F-H+{\bf \Omega \cdot L})/k_B T\big)
\end{equation} 
The notion of thermal equilibrium does not appear so simple before the 
transformation from the laboratory frame (where $H(t)$ depends on time) 
to the rotating frame (where ${\cal H}$ does not depend on time). It is 
important to realize that Eq.(5) represents an entirely rigorous 
Hamiltonian which describes exactly why a rotating bucket at angular 
velocity ${\bf \Omega}$ induces vorticity and net angular momentum into 
the fluid contained within the bucket. 

The transition from a laboratory frame TOP trap time varying Hamiltonian 
$H(t)$ to a rotating frame {\em time independent Hamiltonian} 
${\cal H}$ is a bit more subtle but will now be shown to be closely 
analogous. 

A TOP trap for triplet (spin one) atoms is simply a time dependent magnetic 
field ${\bf B}({\bf r},t)$ constructed by superimposing a homogeneous rotating 
magnetic field ${\bf B}_h(t)$ together with a non-rotating inhomogeneous 
quadrapole magnetic field ${\bf B}_Q({\bf r})$; In detail, 
\begin{equation}
{\bf B}({\bf r},t)={\bf B}_h(t)+{\bf B}_Q({\bf r}),
\end{equation}
where the quadrapole field is given in terms of the field gradient 
amplitude $G$, 
\begin{equation}
{\bf B}_Q({\bf r})=G\big({\bf r}-3({\bf n\cdot r}){\bf n}\big).
\end{equation}
The homogeneous field ${\bf B}_h$ rotates about (and is normal  
to) the rotation axis unit vector ${\bf n}$; In detail, 
${\bf n\cdot B}_0=0$ and  
\begin{equation}
{\bf B}_h(t)={\bf B}_0 \cos(\Omega t)+
{\bf n\times B}_0 \sin(\Omega t). 
\end{equation}
A triplet state (spin one) atom interacts with the magnetic field 
${\bf B}$ so that there exists three possible (potential) energy levels, 
namely zero energy and $\pm \hbar \gamma |{\bf B}|$, where 
$\gamma$ denotes the magnitude of the gyro-magnetic ratio. An atom in 
only one of these states can be trapped with the potential  
\begin{equation}
U({\bf r},t)=\hbar \gamma |{\bf B}({\bf r},t)|,
\end{equation}
The trap potential in cylindrical coordinates 
${\bf r}=(\rho ,\varphi ,z)$ (where $z={\bf n\cdot r}$) is determined by  
\begin{equation}
|{\bf B}|=
G\sqrt{\rho^2+b^2+2b\rho \cos(\varphi -\Omega t)+4z^2}.
\end{equation}
The magnetic length of the TOP trap is defined as $b=|{\bf B}_0|/G$.

For $N$ atoms in a TOP trap, the time dependent Hamiltonian has the form 
\begin{equation}
H(t)=\sum_{1\le j\le N} h_j(t)+\sum_{1\le j\le k\le N}v_{jk}.
\end{equation}
The two body pair interaction on the right hand side of Eq.(12) is  
assumed to conserve total angular momentum. but the time dependent 
single particle Hamiltonian sum on the right hand side of Eq.(12) {\em 
does not conserve total angular momentum} ${\bf n\cdot L}$; i.e.   
\begin{equation}
h_j(t)=-\Big({\hbar^2\over 2M}\Big)\nabla_j^2 +
U(\rho_j ,\varphi_j -\Omega t,z_j) 
\end{equation} 
has a potential which depends on the angle $\varphi_j$, as well as 
$\rho_j$ and $z_j$. 

One may now solve the Schr\"odinger equation for the many body wave 
function $\chi$, 
\begin{equation}
i\hbar \Big({\partial \chi (...,{\bf r}_j,...,t)\over \partial t}\Big)=
H(t)\chi (...,{\bf r}_j,...,t),
\end{equation}
by looking for a solution of the form 
\begin{equation}
\chi (...,\rho_j,\varphi_j,z_j...,t)=
\Psi (...,\rho_j,\varphi_j-\Omega t,z_j...,t)
\end{equation}
yielding 
\begin{equation}
i\hbar \Big({\partial \Psi (...,{\bf r}_j,...,t)\over \partial t}\Big)=
{\cal H}\Psi (...,{\bf r}_j,...,t).
\end{equation}
The new Hamiltonian ${\cal H}$ {\em does not depend on time},
\begin{equation}
{\cal H}=\sum_{1\le j\le N} \tilde{h}_j+
\sum_{1\le j\le k\le N}v_{jk},
\end{equation}
\begin{equation}
\tilde{h}_j=-\Big({\hbar^2\over 2M}\Big)\nabla_j^2 +
U({\bf r}_j)
+i\hbar \Omega \Big({\partial \over \partial \varphi_j}\Big).
\end{equation}
Eqs.(17) and (18) yield the Hamiltonian form  
\begin{equation}
{\cal H}({\rm TOP\ trap})=H-{\bf \Omega \cdot L}
=H-\Omega L_z
\end{equation}
which {\em does not depend on time}, {\em does not conserve 
angular momentum} but does represent the central result of 
this work. In the rotating frame of the TOP trap, the single atom 
potential does not depend on time
\begin{equation}
U({\bf r})=\hbar \gamma G
\sqrt{\rho^2+b^2+2b\rho \cos\varphi + 4z^2},
\end{equation}
and does not conserve the angular momentum component 
$L_z={\bf n\cdot L}$. 

In the previous theoretical literature concerning Bose condensates in 
TOP traps\cite{9,10}, a time averaged Hamiltonian was employed  
\begin{equation}
\big<H\big>=\sum_{1\le j\le N} \big<h_j\big>
+\sum_{1\le j\le k\le N}v_{jk},
\end{equation}
where 
\begin{equation}
 \big<h_j\big>=-\Big({\hbar^2\over 2M}\Big)\nabla_j^2 +
\bar{U}(\rho_j,z_j)
\end{equation}
with the time averaged potential 
\begin{equation}
\bar{U}(\rho ,z)={\Omega \over 2\pi } \int_0^{2\pi /\Omega }
U(\rho ,\varphi-\Omega t,z)dt.
\end{equation}
We note in passing that the time averaged TOP potential 
$\bar{U}(\rho ,z)$ is not very well approximated by a simple 
anisotropic harmonic oscillator potential. More to our 
central point, there is no justification for employing the time 
averaged Hamiltonian $\big<H\big>$ for atoms in a TOP trap. 
Eqs.(17), (18) and (20) provide (in a mathematically rigorous fashion) 
the appropriate Hamiltonian ${\cal H}$ for a TOP trap, just as 
the general Eq.(5) has long provided the appropriate Hamiltonian for 
superfluid $He^4$ in a rotating bucket.\cite{11} 

In the Gross-Piteavskii\cite{12,13} dilute quantum gas model of a 
Bose condensate, the TOP trap induced order parameter of the condensate  
ought to obey the an equilibrium equation which follows from our above  
considerations; It is     
\[
\Big\{-{\hbar ^2\over 2M}\Big({\partial^2\over \partial z^2}+
{\partial^2 \over \partial \rho^2}+
{1\over \rho}{\partial \over \partial \rho}+{1\over \rho^2}
{\partial^2 \over \partial \varphi^2}\Big)+
i\hbar \Omega {\partial \over \partial \varphi }
\]
\[
+\ \hbar \gamma G\sqrt{\rho^2+b^2+2b\rho \cos\varphi + 4z^2}
\ \Big\}\psi (\rho ,\varphi ,z)\ +
\]
\begin{equation}
\Big({4\pi \hbar^2 a |\psi (\rho ,\varphi ,z)|^2\over M}\Big)
\psi (\rho ,\varphi ,z)=
\mu \psi (\rho ,\varphi ,z), 
\end{equation}
where $\mu $ is the chemical potential, and $a$ is the two body 
scattering length. The ground state order parameter Eq.(24) is more 
than a little complicated. However, some of the implications of 
Eq.(24) may be stated with confidence. 

Only in the limit $\Omega \to 0$ can the Bose condensate order 
parameter $\psi ({\bf r})$ be chosen to be real. For the TOP trap 
situation with $\Omega \ne 0$, the order parameter is complex 
\begin{equation}
\psi ({\bf r})=\sqrt{n_s({\bf r})}\exp\big(iM\Phi ({\bf r})/\hbar)\big)
\end{equation}
yielding via Eqs.(24) and (25) a non-trivial superfluid flow velocity 
${\bf v}_s={\bf \nabla }\Phi $ and a non-zero mean angular momentum. 
The velocity flow is determined by a rotational quantum Bernoulli 
equation 
\begin{equation}
\mu=M\Big({1\over 2}|{\bf \nabla}\Phi|^2-({\bf \Omega \times r})
{\bf \cdot \nabla}\Phi \Big)+\tilde{U},
\end{equation}
where 
\begin{equation}
\tilde{U}=U-{\hbar ^2\over 2M}\Big({\nabla^2 \sqrt{n_s}\over 
\sqrt{n_s}}\Big)+\Big({4\pi \hbar^2 a n_s\over M}\Big).
\end{equation}

If the TOP trap were not rotating, the small scattering length of two 
$^{87}$Rb atoms ($a\sim 50$\AA ), along with the small density of the  
condensate\cite{4} $n_s$, would lead to a localization length within 
the trap potential\cite{14} minimum determined by 
$d\sim (\hbar /M\gamma G)^{1/3}\sim .15\mu m$. 
This localization length $d$ is much smaller than the magnetic 
length of $b\sim 8\times  10^2 \mu m$, i.e. $d<<b$. 
If the condensate stayed within a distance $d$ of the minimum of 
the potential $U({\bf r})$, then there would be frequent spin 
flips resulting in the loss of atoms from the trap. The rotating potential 
causes the condensate atoms climb a bit up the potential wall away 
from the center, acting in a manner closely analogous to particles in a 
centrifuge. Thus lifted from the potential minimum, the spin induced 
lifetimes of atoms in the trap are increased.

To see how the mechanics of the rotational equilibrium for the condensate 
density $n_s$ may work, let us first make an analogy with a roulette wheel 
and a rolling metal ball found in many gambling casinos (not yet having 
completely entered the digital computer game simulation age). The mechanical 
roulette wheel turns and the ball is rolled into a grove of radius 
$b$. The ball rolls around this grove while the wheel spins. Along the 
grove, there are many minor potential minima, i.e. one minor minimum 
for every possible number that can be bet to win on the roulette wheel. 
The wheel continues to turn, and finally the ball 
falls into one of the potential minima. The ball then rotates with the 
same angular velocity as the roulette wheel achieving ``rotational thermal 
equilibrium'' at the winning number.  The point 
is that the rotational velocity of the  ball relative to the rotational 
velocity of the roulette wheel cannot differ from zero for a very long 
time. Otherwise, the bets would never get settled. The ball does 
{\em not} in the final stages of the game feel the 
``{\bf T}ime average {\bf O}f the {\bf P}otential'', i.e that time 
average in which the ball is continually bouncing over the minor 
potential barriers  during the initial stages of the game. 
The ball does {\em finally settle 
down to a smooth rotational free energy minimum}. 

In the TOP trap, the condensate is placed in a rotating potential grove 
of radius $b$. The Bose condensate can hardly stand still 
for very long while feeling only the ``Time average Of the Potential''; 
i.e. the condensate can only shake rapidly up and down the $z$-axis for 
a limited amount of time. Then the condensate then starts to flow smoothly 
around the grove. The critical angular velocity for forming one quantum 
of circulation,  
\[
\kappa_0=(2\pi \hbar /M)\sim 4.6\times 10^{-5} cm^2/sec 
\]
around to rotation axis cannot be very much larger than   
$\Omega_c\sim (\kappa_0 /b^2)$ which is (in order of magnitude) how slowly 
the minute hand of a clock rotates. 
This is a much lower rotational velocity then the TOP angular 
velocity\cite{13} of $(\Omega /2\pi)\sim 7.5\times 10^3 Hz$. One thus 
expects perhaps $\sim 10^5$ circulation quanta to flow around the axis 
of rotation. This circulation represents a very large number of quanta indeed. 
Unlike the roulette wheel there is but one potential minima per turn around 
the grove. The circulating Bose condensate, then flows up and down the 
potential as it forms in its own little oval flow (as in a toroidal pipe) 
around this funnel potential. 

What happens to this flow of condensate around a toroidal pipe 
(so to speak) when the TOP trap is removed? The diameter of the pipe 
will increase and the tangential velocities to the flow around the ring 
will decrease (due to angular momentum conservation). The oval which is 
left, and displayed in an experimental picture could hardly be circular. 
It would be elliptic due to the displaced asymmetric axis of rotation. 
The above considerations are consistent with TOP trap experiments.

\end{document}